\documentclass[preprint]{aastex}
\usepackage{times}
\usepackage{graphicx}

\newcommand\phtneg{\phantom{$-$}}
\newcommand\phtdig{\phantom{2}}
\newcommand\phtast{\phantom{*}}

\shorttitle{The Lunokhod~1 Reflector}
\shortauthors{Murphy et al.}

\begin{document}

\title{Laser Ranging to the Lost Lunokhod~1 Reflector}

\author{
T.\,W. Murphy,~Jr.\altaffilmark{1},
E.\,G. Adelberger\altaffilmark{2},
J.\,B.\,R. Battat\altaffilmark{3},
C.\,D. Hoyle\altaffilmark{4},
N.\,H. Johnson\altaffilmark{1},
R.\,J. McMillan\altaffilmark{5},
E.\,L. Michelsen\altaffilmark{1},
C.\,W. Stubbs\altaffilmark{6},
H.\,E. Swanson\altaffilmark{2}}
\email{tmurphy@physics.ucsd.edu}

\altaffiltext{1}{University of California, San Diego, Dept. of Physics, La Jolla, CA 92093-0424}
\altaffiltext{2}{University Washington, Dept. of Physics, Seattle, WA 98195-1560}
\altaffiltext{3}{Massachusetts Institute of Technology, Dept. of Physics, Cambridge, MA 02139}
\altaffiltext{4}{Humboldt State University, Dept. of Physics and Astronomy, Arcata, CA 95521-8299}
\altaffiltext{5}{Apache Point Observatory, Sunspot, NM 88349-0059}
\altaffiltext{6}{Harvard University, Dept. of Physics, Cambridge, MA 02318}

\begin{abstract}
In 1970, the Soviet Lunokhod~1 rover delivered a French-built laser
reflector to the Moon. Although a few range measurements were made
within three months of its landing, these measurements---and any that
may have followed---are unpublished and unavailable. The Lunokhod~1
reflector was, therefore, effectively lost until March of 2010 when
images from the Lunar Reconnaissance Orbiter (LRO) provided a positive
identification of the rover and determined its coordinates with uncertainties
of about 100~m. This allowed the Apache Point Observatory Lunar Laser-ranging
Operation (APOLLO) to quickly acquire a laser signal. The reflector
appears to be in excellent condition, delivering a signal roughly
four times stronger than its twin reflector on the Lunokhod~2 rover.
The Lunokhod~1 reflector is especially valuable for science because
it is closer to the Moon's limb than any of the other reflectors and,
unlike the Lunokhod~2 reflector, we find that it is usable during
the lunar day. We report the selenographic position of the reflector
to few-centimeter accuracy, comment on the health of the reflector,
and illustrate the value of this reflector for achieving science goals. 
\end{abstract}


\section{Introduction}

The Apollo~11 astronauts that landed on the Moon in July, 1969 deployed
a retro-reflector intended for lunar laser ranging (LLR). Shortly
thereafter, several ground stations, including the Lunar Ranging Experiment
(LuRE) at the McDonald Observatory 2.7~m telescope, ranged to the
reflector. In November, 1970, the Soviets landed the Lunokhod~1 robotic
rover that carried a French-built reflector with a different design
but a comparable expected reflection efficiency. The Soviets \citep{kokurin}
and the French \citep{calame} both ranged to the reflector during
its first lunar night in December of 1970, followed by another success
in February, 1971. The LuRE team tried to locate
the Lunokhod~1 reflector, and may have seen glimpses of it a time
or two, but never enough to pin down its position. Kokurin relates
in 1975 that the Lunokhod~1 reflector was seen again in May of 1974,
but no details are provided, and the data are not available. The Lunokhod~1
reflector was followed by the the Apollo~14, Apollo~15, and Lunokhod~2
reflectors, arriving in that order. We denote these reflectors as
A11, L1, A14, A15, and L2, respectively. The arrival of other reflectors
whose coordinates were well known, and which in the case of A15 had
much larger effective area, reduced the motivation for spending valuable
telescope time on continued searches for L1.

The position of the Lunokhod~1 rover was known to within approximately
5~km \citep{stooke}. This large uncertainty was a problem for LLR,
which typically uses a narrow time window to reduce the inevitable
background. For example, the Apache Point Observatory Lunar Laser-ranging
Operation \citep[APOLLO:][]{apollo} normally employs a 100~ns detector
gate, translating to a one-way range window with a depth of 15~m.
Because the lunar surface normal at L1 is angled 50$^{\circ}$ to
the line of sight from Earth, the range window maps to a 20~m wide band
on the lunar surface. Positional uncertainty of a few kilometers translates
into a vast search space so that detecting the reflector is unlikely.
Nonetheless, APOLLO spent a small fraction of its telescope time on
favorable nights unsuccessfully exploring the space around the best-guess
coordinates of \citet{stooke}.

A breakthrough occurred in 2010 March, when the Lunar Reconnaissance
Orbiter Camera \citep[LROC:][]{lroc} obtained a high-resolution image of
the Luna~17 landing area that clearly revealed the lander, rover, and rover
tracks. The coordinates obtained by the LROC team had an uncertainty of
about 100~m \citep{plescia} and were centered on a region almost 5~km from
the previous best estimate. Additionally, laser altimetry from the Lunar
Orbiter Laser Altimeter \citep[LOLA:][]{lola} on board LRO determined the
site radius to better than 5~m \citep{neumann}.  But it was not possible to
know if L1 was correctly oriented toward the Earth. A few unsuccessful
ranging attempts were made in late March in unfavorable observing
conditions.  The first available favorable observing time on 2010
April 22 yielded immediate results, revealing a return that appeared 270~ns
later than our prediction.  The initial return was surprisingly bright, far
surpassing the best-ever return signal from the twin reflector on
Lunokhod~2.

\section{Observations}

Following the initial discovery on April 22, we continued to
successfully range to L1 even when seeing conditions were
not optimal or when the reflector was illuminated by the sun. We were
surprised that L1 remained visible during lunar daylight because,
unlike the APOLLO reflectors that are usable in lunar daylight, L2
(L1's twin) cannot be seen when the sun shines on the reflector, even
in excellent observing conditions. %
\begin{table}
\begin{center}
\caption{Observation summary\label{tab:obs}}
\small
\begin{tabular}{lcccccccc}
Date&
 Phase, $D$&
L1 sun elevation&
 Libration (lon,lat)&
 A11&
L1&
 A14&
 A15&
 L2\tabularnewline
\hline
2010-04-22&
 96$^{\circ}$&
 $-20^{\circ}$&
 ($-2.9^\circ$, $2.7^\circ$)&
 1&
 4&
 1&
 2&
 ---\tabularnewline
2010-04-26&
 150$^{\circ}$&
 \phtneg 18$^{\circ}$&
 ($2.1^\circ$, $7.3^\circ$)&
 2&
 1&
 2&
 4&
 --\tabularnewline
2010-05-05&
 263$^{\circ}$&
 \phtneg 33$^{\circ}$&
 ($1.9^\circ$, $-1.9^\circ$)&
 1&
 2&
 2&
 3&
 2\tabularnewline
2010-05-23&
 118$^{\circ}$&
 \phtdig $-7^{\circ}$&
 ($2.5^\circ$, $7.1^\circ$)&
 1&
 1&
 1&
 4&
 ---\tabularnewline
2010-05-24&
 132$^{\circ}$&
 \phtneg\phtdig 3$^{\circ}$&
 ($3.3^\circ$, $7.4^\circ$)&
 1&
 2&
 1&
 9&
 ---\tabularnewline
2010-06-16&
 49$^{\circ}$&
 $-49^{\circ}$&
 ($-0.8^\circ$, $4.2^\circ$)&
 2&
 3&
 2&
 3&
 2\tabularnewline
2010-06-20&
 102$^{\circ}$&
 $-21^{\circ}$&
 ($4.2^\circ$, $7.5^\circ$)&
 1&
 2&
 2&
 2&
 ---  \tabularnewline
2010-08-18&
106$^{\circ}$&
$-21^{\circ}$&
($7.5^\circ$, $4.1^\circ$)&
1&
2&
2&
3&
---\tabularnewline
\hline
\end{tabular}\end{center}
\end{table}

Table~\ref{tab:obs} summarizes our L1 observations, also enumerating range
measurements to the other reflectors used to determine the lunar orientation.
All sessions, except for 2010 May 24, lasted less than one hour. The
libration offsets in Table~\ref{tab:obs} incorporate topocentric
corrections based on the viewing angle from the Apache Point Observatory
(APO)
at the time of observation, in effect representing the longitude and
latitude of the point on the Moon directly ``below'' APO.  These viewing angles establish the geometry
relevant to constraining the position of L1 on the moon.  Given the
location of L1 on the Moon, the best constraints would arise from maximal
librations in the first and third quadrants, i.e., roughly $(7^\circ ,\
7^\circ )$, and $(-7^\circ ,\ -7^\circ )$.  In this case, no observations
occupy the third quadrant, and many cluster around similar values.  As a
result, the centimeter-level position constraints presented here will
improve substantially as more observations are made over a broader range of
viewing angles.

\section{Location of the L1 Reflector}

During the first night of observation, the viewing geometry changed
enough in 40 minutes to place a weak constraint on the position of
the reflector. The error ellipse had an aspect ratio of almost 1000:1,
amounting to almost a linear constraint roughly 50 meters long. After
2010 May 5 we could fit for a position accurate to approximately one
meter.

Using the L1 range points listed in Table~\ref{tab:resids} we solved for
the coordinates of the L1 reflector. Before the range residuals---the
difference between the observation and a prediction based on the Jet
Propulsion Laboratory's DE421 ephemeris---can be used to solve for the
reflector position, adjustments for lunar tidal deformation and for Earth
and lunar orientations (librations) must be made.  Table~\ref{tab:resids}
does not include the initial observations from 2010 April 22 because these
observations used a 600~ns detector gate to aid acquisition, rather than
the nominal well-calibrated 100~ns gate.

For consistency with DE421, the tidal model assumed second-degree
Love numbers $h_{2}=0.0379$ and $l_{2}=0.0105$, so that the tidal
displacement vector for a point in the direction $\hat{r}$ from the
center of the Moon is\[
\Delta\vec{r}=H\left(\frac{R_{0}}{R}\right)^{3}\left\{ h_{2}\hat{r}\left[\frac{3}{2}\left(\hat{R}\cdot\hat{r}\right)^{2}-\frac{1}{2}\right]+3l_{2}\left(\hat{R}\cdot\hat{r}\right)\left[\hat{R}-\left(\hat{R}\cdot\hat{r}\right)\hat{r}\right]\right\} ,\]
where $\hat{R}$ is the unit vector from Moon center to the source
of the tidal deformation (Earth or Sun), $R$ is the center--to--center
distance, $R_{0}$ is the time-average distance, and $H$ is the displacement
parameter. For Earth-induced tides, $H=12.89$~m, and $H=0.074$~m
for solar tides. Given the small value for the Love numbers, the tidal
amplitude from the Earth is reduced to 0.489~m, and to 2.8~mm for
the sun. We applied range offsets to each observation corresponding
to the line-of-sight projection of the tidal displacement at the time
of observation. Rather than correct to the zero-tide lunar figure,
we corrected to the mean, or ``static'' tidal figure of the tidally-locked
Moon, so that corrections were typically less than 0.1~m.

The prediction used in this analysis employs a simple Earth orientation
model based on extrapolations of the most recent Earth orientation
parameters at the time of observation, resulting in range offsets of
several nanoseconds and drift rates approaching 1~ns per hour. Rather than
use the best available post-fit Earth orientation parameters, which are
accurate to several millimeters, we empirically null Earth orientation
effects by subtracting the offset and trend of residuals to the Apollo~15
reflector.  The Apollo~15 observations typically bracket observations of
the other reflectors and have the most range measurements at the highest
precision.  This ``trend correction'' produced the residuals shown in the
fifth column of Table~\ref{tab:resids}. Residuals in Table~\ref{tab:resids}
are referenced to the best-fit L1 coordinates. The number of photons used
in the reduction represent a subset of the detected photons.

The residuals from the different reflectors differed systematically,
indicating errors in the lunar orientation/libration as derived from
the DE421 ephemeris. We optimized the lunar orientation by minimizing
the residuals of the various reflectors, excluding L1 whose position
we were trying to determine. Measurements of the three Apollo reflectors
determine the mean range and two orientation angles (the data are
insensitive to lunar rotations about the line from the center of the
Moon to the observatory). When L2 data are added, the rigid-body fit
is over-determined and is found by least-squares minimization, appropriately
weighting data points by their measurement uncertainties. The sixth
column in Table~\ref{tab:resids} displays residuals after the lunar
orientation corrections are applied to the L1 ranges. In all cases,
the orientation corrections were less than 0.15~m on the lunar surface,
projecting to less than this in range.

In fitting for an improved reflector position, we evaluated the partial
derivatives of the ranges---corrected for tides and orientation---with
respect to changes in radius, latitude, and longitude of the reflector
from the approximate position determined after 2010 May 5. The partial derivatives
were evaluated for each measurement, using the topocentric-corrected
libration angle at that time. A least-squares method minimized the
chi-squared values of the difference between the measured residual
and the predicted location-adjusted range residual from the partial
derivatives.%
\begin{table}
\begin{center}
\caption{L1 range residuals and uncertainties.\label{tab:resids}}
\small
\begin{tabular}{lccccc}
Date&
 Observation \#&
 \# photons&
 1$\sigma$ error&
\multicolumn{2}{c}{residual (mm)}\tabularnewline
&
&
 used&
 (mm)&
 trend-corrected&
 libration-corrected\tabularnewline
\hline
2010-04-26&
 1&
 12&
 19.3&
 $-6.1$&
 $-34.4$\tabularnewline
2010-05-05&
 2&
 149&
 3.6&
29.0&
 \phtdig $-3.0$\tabularnewline
&
 3&
 42&
 6.9&
44.8&
 \phtneg 12.8\tabularnewline
2010-05-23&
 4&
 283&
 2.6&
 30.3&
 \phtneg\phtdig 6.2\tabularnewline
2010-05-24&
 5&
 23&
 5.4&
 32.1&
 \phtdig $-4.0$\tabularnewline
&
 6&
 90&
 5.0&
 29.4&
 \phtdig $-6.7$\tabularnewline
2010-06-16&
 7&
 104&
 3.7&
 21.2&
\phtneg\phtdig 0.0 \tabularnewline
&
 8&
 45&
 5.3&
16.3&
 \phtdig $-4.9$\tabularnewline
&
 9&
 55&
 5.4&
 19.9&
 \phtdig $-1.2$\tabularnewline
2010-06-20&
 10&
 76&
 4.5&
 30.5&
\phtdig $-2.4$ \tabularnewline
&
 11&
 14&
 8.6&
 23.9&
 \phtdig $-9.0$ \tabularnewline
2010-08-18&
12&
409&
2.5&
21.6&
\phtneg\phtdig 3.7\tabularnewline
&
13&
224&
2.9&
12.9&
\phtdig $-5.0$\tabularnewline
\hline
\end{tabular}\end{center}
\end{table}

\begin{figure}
\begin{center}\includegraphics[%
  scale=0.7]{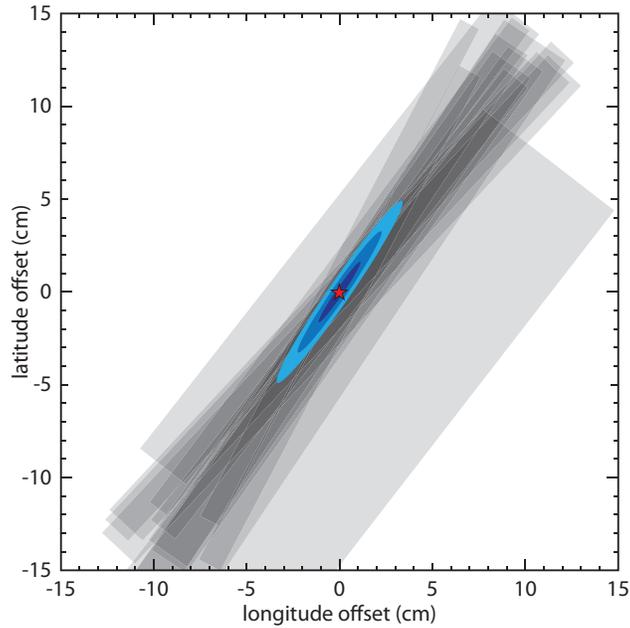}\end{center}

\caption{Coordinate constraints for the 13 observations in Table~\ref{tab:resids},
confined to the surface at $r=1734928.72$~m. Gray bands represent
individual 1-$\sigma$ constraints, darkening when they overlap, and
truncated in length to facilitate individual identification. The joint
probability is indicated by the error ellipses, at 1-$\sigma$, 2-$\sigma$,
and 3-$\sigma$ levels. \label{fig:error-ellipse}}
\end{figure}

Using all 13 observations in Table~\ref{tab:resids}, we find a position
solution with one-sigma errors in the east, north and vertical directions of
$(0.026,\ 0.030,\ 0.025)$~meters, respectively.
In accordance with the usual practice \citep{pdg}, we accounted
for the quality of the fit, ($\chi^{2}/\nu=2.32$ for $\nu=10$ degrees
of freedom), by multiplying the 1-$\sigma$ formal errors by $\sqrt{\chi^{2}/\nu}$.
Figure~\ref{fig:error-ellipse} shows the observational constraints
on latitude and longitude for a fixed radius.

Table~\ref{tab:l1-coords} shows our best-fit L1 coordinates referred to the
Jet Propulsion Laboratory's DE421 ephemeris coordinate system for the Moon.
Both principal-axis and mean-Earth coordinates are provided, in spherical
and Cartesian coordinates.  For spherical coordinates, $\phi$ refers to
latitude and $\lambda$ refers to longitude.  The static part of the lunar
tide is incorporated in the coordinates for more accurate predictions in
the absence of a lunar tidal model. The $X$-component of the Cartesian set
is especially well-determined because this is principally along the
direction of the range measurements.%
\begin{table}
\begin{center}
\caption{L1 Position in DE421 coordinates\label{tab:l1-coords}}
\small
\begin{tabular}{lccc}
\hline
Principal Axis, ($r$, $\phi$, $\lambda$)&
 1734928.72$\pm0.03$&
 38.3330784$\pm0.000001$&
 $-35.036674\pm0.000001$\tabularnewline
Principal Axis, ($X$, $Y$, $Z$)&
 1114292.387$\pm0.004$&
 $-781299.33\pm0.03$&
 1076058.31$\pm0.03$\tabularnewline
Mean Earth, ($r$, $\phi$, $\lambda$)&
 1734928.72$\pm0.03$&
 38.3151577$\pm0.000001$&
 $-35.007964\pm0.000001$\tabularnewline
Mean Earth, ($X$, $Y$, $Z$)&
 1114959.354$\pm0.004$&
 $-780933.94\pm0.03$&
 1075632.60$\pm0.03$ \tabularnewline
\hline
\end{tabular}\end{center}
\end{table}

Table~\ref{tab:all-5} includes the coordinates for all five reflectors in
the DE421 principal-axis system, repeating coordinates for L1 found in Table~\ref{tab:l1-coords}.  The static
piece of the lunar tide, given in Table~\ref{tab:tides}, is again
incorporated in the coordinates. To obtain the reflector coordinates on the
un-deformed Moon, subtract the values in Table~\ref{tab:tides} from those
in Table~\ref{tab:all-5}.

\begin{table}
\begin{center}
\caption{Coordinates of the five reflectors, in the DE421 principal-axis frame\label{tab:all-5}}
\footnotesize
\begin{tabular}{lrrrrrr}
Reflector&
 \multicolumn{1}{c}{$r$ (m)}&
 \multicolumn{1}{c}{$\phi$ (deg)}&
 \multicolumn{1}{c}{$\lambda$ (deg)}&
 \multicolumn{1}{c}{$X$ (m)}&
 \multicolumn{1}{c}{$Y$ (m)}&
 \multicolumn{1}{c}{$Z$ (m)}\tabularnewline
\hline
Apollo 11&
 1735473.105&
  0.6934308&
  23.4543026&
 1591967.923&
  690698.118&
  21003.312\tabularnewline
Lunokhod 1&
 1734928.72\phtdig &
  38.3330784&
 $-35.036674$\phtdig &
 1114292.387&
 $-781299.33$\phtdig &
  1076058.31\phtdig \tabularnewline
Apollo 14&
 1736336.555&
  $-3.6233280$&
 $-17.4971027$&
 1652689.795&
 $-520999.212$&
 $-109731.020$\tabularnewline
Apollo 15&
 1735477.684&
  26.1551690&
  3.6103512&
 1554679.329&
  98094.120&
  765004.914\tabularnewline
Lunokhod 2&
 1734639.201&
 \phtneg 25.8509889&
 \phtneg 30.9087373&
 1339364.890&
 \phtneg 801870.780&
 \phtneg 756358.447  \tabularnewline
\hline
\end{tabular}\end{center}
\end{table}

\begin{table}
\begin{center}
\caption{Static tidal offsets for the five reflectors\label{tab:tides}}
\small
\begin{tabular}{lccc}
Reflector&
$\Delta r$ (m)&
$r\Delta\phi$ (m)&
$r\Delta\lambda\cos\phi$ (m)\tabularnewline
\hline
Apollo 11&
$0.3724$&
$-0.0041$&
$-0.1482$\tabularnewline
Lunokhod 1&
$0.0580$&
$-0.1324$&
\phtneg $0.1497$\tabularnewline
Apollo 14&
$0.4196$&
\phtneg $0.0233$&
\phtneg $0.1162$\tabularnewline
Apollo 15&
$0.3438$&
$-0.1600$&
$-0.0229$\tabularnewline
Lunokhod 2&
$0.1926$&
$-0.1173$&
$-0.1610$\tabularnewline
\hline
\end{tabular}\end{center}
\end{table}

\section{Reflector Performance}

We observed a surprisingly strong signal on APOLLO's 2010 April 22
discovery of L1, during which we concentrated on verifying the signal
by adjusting timing parameters rather than optimizing the signal via
pointing adjustments. Nevertheless, we detected 1916 photons from
L1 in a 10,000 shot run, compared to 2040 photons from A15 in a 5000
shot run. This suggested that L1 could be expected to deliver at least
half the signal strength of the large A15 reflector.

\begin{table}
\begin{center}
\caption{Normalized return rates for the five reflectors; asterisks denote
dark conditions\label{tab:rates}}
\small
\begin{tabular}{lccccccc}
Date&
L1 sun elevation&
 Rate Factor&
 A11&
L1&
 A14&
 A15&
 L2\tabularnewline
\hline
2010-04-22&
 $-20^{\circ}$&
 0.0894&
 0.47\phtast &
 1.70{*}&
 1.23{*}&
 3.30\phtast &
 ---\tabularnewline
2010-04-26&
 \phtneg 18$^{\circ}$&
 0.0383&
 0.08\phtast &
 0.06\phtast &
 1.24\phtast &
 3.76\phtast &
 ---\tabularnewline
2010-05-05&
 \phtneg 33$^{\circ}$&
 0.1770&
 0.61{*}&
 0.23\phtast &
 1.06\phtast &
 3.33\phtast &
 0.17{*}\tabularnewline
2010-05-23&
 \phtdig $-7^{\circ}$&
 0.0864&
 1.12\phtast &
 0.79{*}&
 0.56\phtast &
 3.32\phtast &
 ---\tabularnewline
2010-05-24&
 \phtneg\phtdig 3$^{\circ}$&
 0.0116&
 1.19\phtast &
 1.08\phtast &
 0.61\phtast &
 3.20\phtast &
 ---\tabularnewline
2010-06-16&
 $-49^{\circ}$&
 0.0426&
 0.96{*}&
 0.59{*}&
 1.08{*}&
 2.96{*}&
 0.14{*}\tabularnewline
2010-06-20&
 $-21^{\circ}$&
 0.0077&
 2.34\phtast &
 1.49{*}&
 0.78{*}&
 1.88\phtast &
 ---  \tabularnewline
2010-08-18&
$-21^{\circ}$&
0.0410&
0.74\phtast &
1.80{*}&
0.77{*}&
3.49\phtast &
--- \tabularnewline
\hline
\end{tabular}\end{center}
\end{table}

Table~\ref{tab:rates} compares the normalized observed return rates
for the different reflectors. The normalization factor was derived
by adding the signal strengths, in photons per shot, for the three
Apollo reflectors for the night, and dividing by five. Because the
Apollo reflectors have cross-sections in a 1:1:3 ratio, this effectively
normalizes the rate to a standard A11/A14 reflector. We find that
L1 is approximately four times as bright as L2 when both reflectors
are in the dark, and when L1 is illuminated its brightness is comparable
to that of L2 in the dark. Table~\ref{tab:accum-rates} summarizes
the results. The Apollo reflectors behave, on average, in their expected
1:1:3 ratio. Averaged over a lunar day, L1 performs comparably to
the smaller Apollo arrays, and somewhat better when L1 is in the dark.
Our preliminary results are very encouraging for the potential contribution
L1 will make to LLR science.

\begin{table}
\begin{center}
\caption{Accumulated return rate statistics\label{tab:accum-rates}}
\small
\begin{tabular}{lccccc}
Condition&
 A11&
L1&
 A14&
 A15&
 L2\tabularnewline
\hline
All&
 0.94&
 0.97&
 0.92&
 3.16&
 0.155\tabularnewline
L1 dark&
 ---&
 1.27&
 ---&
 ---&
 ---\tabularnewline
L1 daylight&
 ---&
 0.46&
 ---&
 ---&
 ---  \tabularnewline
\hline
\end{tabular}\end{center}
\end{table}

The difference in L1's and L2's performance is striking, and as yet
unexplained. The LuRE group found that L2 was initially 25\% brighter
than the A15 reflector. Today, APOLLO typically sees L2 perform at
one-tenth the A15 rate. The Apollo reflectors have degraded by as
much as a factor of ten since their installation \citep{degrade}, and
L2 has clearly degraded more than the Apollo arrays. The fact that
L1 performs comparably to the Apollo arrays puts the spotlight on
L2, whose rate of degradation appears to be anomalous.

The relative expected brightness of the pristine Lunokhod and Apollo
reflectors can be calculated as follows. At normal incidence, the
triangular-faced Lunokhod corner cubes---measuring 11~cm on a side---have
a hexagonal reflecting area of 3493~mm$^{2}$. The 3.8~cm diameter
Apollo cubes have a reflective area of 1134~mm$^{2}$ at normal incidence.
But the peak intensity of the far-field diffraction pattern from an
individual cube scales approximately as the area of the reflector
squared. Moreover, the central intensity of an uncoated corner cube
relying on total internal reflection---like the Apollo cubes---is
only 27\% that of a perfect Airy pattern as delivered by a perfectly-coated
corner cube. On balance, the central intensity of a single Lunokhod
cube will therefore be about 30 times that of an Apollo cube if the
reflective coatings on the Lunokhod reflectors operate at 95\% efficiency.
Taking into account the multiple reflectors---14 in a Lunokhod array
vs. 100 in the A11/A14 arrays---one expects the Lunokhod central intensity
to be 4.2 times stronger than that from the A11/A14 arrays.

However, the transverse motion of the Moon together with Earth rotation
typically impose a velocity aberration on the return of $\delta=4$--6~$\mu$rad,
so that the telescope does not sample the central intensity of the
returning far-field diffraction pattern. Because the diffraction pattern
from the Lunokhod corner cubes is significantly narrower than that
from the Apollo cubes, the Lunokhod cubes suffer a greater loss from
velocity aberration. Establishing an effective diameter, $D_{\mathrm{eff}}=66.7$~mm,
for the Lunokhod cubes so that $\frac{\pi}{4}D_{\mathrm{eff}}^{2}=3493\,\mathrm{mm^{2}}$,
we find that the Airy pattern intensity at an angle, $\delta$, away
from the center is $4J_{1}^{2}(x)/x^{2}$, where $x=\delta\lambda/D_{\mathrm{eff}}$,
and $J_{1}(x)$ is the Bessel function of the first kind, order one.
At a wavelength $\lambda=532$~nm, the velocity aberration results
in an intensity of 0.2--0.52 times the central intensity, with a typical
value (for $\delta=5$~$\mu$rad) of 0.35. Under the same conditions,
the Apollo cubes produce an offset return intensity of 0.62--0.81
times the central intensity, with a typical value of 0.72. As a result,
the ratio of effective strengths for Lunokhod vs. A11/A14 is only
2.0 rather than 4.2. For 532~nm light at normal incidence, we expect
the Lunokhod arrays to deliver twice the return of the A11 and A14
arrays. For the ruby laser used in the initial McDonald operation,
the effect is not as severe, so that the expected ratio is 2.55. It
is therefore not surprising that the L2 return strength was initially
seen to be comparable to that from A15, given that A15 is three times
larger than the A11/A14 arrays.

Because the L1 reflector is performing at the expected level relative to
the Apollo reflectors, we can comment on the orientation of the reflector.
An azimuth offset of $40^\circ$ or more is enough to eliminate any signal
return, while an offset of $21^\circ$ reduces the signal to one-half its
optimal value.  It is therefore likely that the rover/reflector are
oriented within $\sim 20^\circ$ of the intended azimuth.  An elevation
angle offset is less likely given the pre-set mechanical mounting angle of
the array, together with the fact that the terrain is locally flat aside
from craters.  The poor performance of L2 cannot easily be attributed to
orientation, as the reflector delivered a very strong initial reposnse at its final position.

\section{Potential Scientific Impact of Ranging to L1}

LLR provides, by determining the shape of the lunar orbit, the most
precise tests of many aspects of gravity, including the strong
equivalence principle, the constancy of Newton's constant, geodetic
precession, gravitomagnetism, and the inverse square law
\citep{nord-icarus,dickey,jgw-96,jgw-latest,gravmag,adelberger-isl}.  By
determining how the Moon's orientation responds to known gravitational
torques, LLR is also a powerful probe of the internal structure of the
Moon \citep{interior}. In both cases, knowing and understanding the
Moon's orientation is vital: we cannot use LLR to locate the Moon's
center of mass without knowing the three-dimensional orientation of the
lunar body. In fact, lunar orientation uncertainties are currently a
limiting factor in converting high-precision ranging data into tests of
fundamental properties of gravity. The orientation of a rigid body is
determined by three angles. However, the angle specifying the Moon's
rotation about the axis from its center to the telescope cannot be probed
by ranging data as the distances to the reflectors are unchanged by this
rotation.  Ranges to at least three reflectors are needed to constrain
the two relevant angles as well as the distance to the center of the
rigid body. The lunar orientation can thus be found whenever the signal
strengths permit ranging to all three Apollo reflectors in a single
session. APOLLO can often do this because the Apollo reflector
efficiencies are roughly independent of the state of solar illumination.
Ranges to additional reflectors of course improve the orientation
constraints, but more importantly, they now constrain models of the
Moon's tidal deformations that are needed to interpret accurately the
ranging data. Ranges to four reflectors provide a single tidal
constraint; ranging to a fifth reflector doubles the number of
independent constraints on the deformation models. However, this argument
underrates the importance of ranging to L1. L2 currently has one tenth
the reflection efficiency of A15. Moreover, L2 is unusable in lunar
daylight, presumably because of thermal gradients within the corner
cubes. Because it is impractical to range near new Moon, L2 is
effectively available only one third of the time.  Fortunately, as seen
in Table~\ref{tab:rates}, L1 is reasonably efficient during lunar
daylight and quite efficient during lunar night, so that APOLLO will
frequently find four available reflectors---and sometimes five.

The most significant impact of our discovery follows from L1's location
on the Moon. A reflector at the center of the visible face of the
Moon contributes only modestly to knowledge of lunar orientation.
Libration moves this central reflector as much as $10^{\circ}$ from
the Earth--Moon line of sight, so that its sensitivity, $\frac{\partial\rho}{\partial\theta}\approx\sin\theta$,
may approach 0.17, where $\theta$ is the angular displacement of
the reflector from the line connecting the ranging station to the
center of the Moon. A reflector at the limb of the Moon, by contrast,
has a sensitivity of 1.0. L1, located approximately $50^{\circ}$
from the Earth--Moon line, is \emph{the closest of the five reflectors}
to the lunar limb, and has an orientation sensitivity of 0.77. Table~\ref{tab:sensitivity}
lists the sensitivity for all five reflectors. The nominal values
correspond to the sensitivities at the average libration of zero in
longitude and latitude. The range indicates extremes for the libration
that puts the reflector closest to and farthest from the sub-Earth
point on the Moon. Both Lunokhod reflectors offer another advantage
over Apollo reflectors because each provides sensitivity to both longitude
and latitude librations, whereas the Apollo arrays tend to be principally
sensitive to just one of the degrees of freedom.

\begin{table}
\begin{center}
\caption{Orientation sensitivity of the reflectors\label{tab:sensitivity}}
\small
\begin{tabular}{lccccc}
Reflector&
 offset (deg)&
 nominal&
 range&
 nominal longitude&
 nominal latitude\tabularnewline
&
&
sensitivity&
&
sensitivity&
sensitivity\tabularnewline
\hline
Apollo 11&
 23.5&
 0.40&
 0.27--0.52&
 0.40&
 0.01\tabularnewline
Lunokhod 1&
 50.0&
 0.77&
 0.65--0.86&
 0.45&
 0.51\tabularnewline
Apollo 14&
 17.9&
 0.31&
 0.17--0.44&
 0.30&
 0.06\tabularnewline
Apollo 15&
 26.4&
 0.44&
 0.33--0.55&
 0.06&
 0.44\tabularnewline
Lunokhod 2&
 39.5&
 0.63&
 0.49--0.76&
 0.46&
 0.37  \tabularnewline
\hline
\end{tabular}\end{center}
\end{table}

The L1 reflector is nominally about twice as sensitive to libration
as any of the Apollo reflectors. Furthermore, it has a high photon
return rate, often greater than of the A11 and A14 reflectors. We
therefore expect L1 to become APOLLO's highest priority reflector
following signal acquisition on the largest of the reflectors, A15.

\section{Conclusion}

We have, for the first time in over three decades, obtained unambiguous
LLR returns from the Lunokhod~1 reflector and determined its position
with an uncertainty of a few centimeters. We found, to our surprise,
that L1 is a much better reflector than its twin L2, so the process
that degraded L2 has not had a big effect on L1. L1's location on
the Moon makes it the most sensitive reflector to Moon's orientation
in latitude and longitude. Because of L1's strategic location and
high reflection efficiency that degrades only modestly when it is
illuminated by the sun, ranging to this reflector will significantly
advance the precision of LLR and the resulting gravitational and lunar
science.

\acknowledgments

T.M. thanks Jim Williams for alerting him to the new LRO images on 2010
March 18, and for providing historical details of L1 ranging efforts.
Eric Silverberg related the history of the attempts to locate L1 by
the McDonald Observatory. APOLLO is supported by the National Science
Foundation (grant PHY-0602507) and the National Aeronautical and Space
Administration (NASA; grant NNX-07AH04G). T.M. is also supported by
the NASA Lunar Science Institute as part of the LUNAR consortium (NNA09DB30A).
Results in this paper are based on observations obtained with the
Apache Point Observatory 3.5~m telescope, which is owned and operated
by the Astrophysical Research Consortium.

\end{document}